\documentclass[10pt,aps.prl,twocolumn,superscriptaddress]{revtex4-1} %% REVTeX 4.0

\usepackage{graphicx}
\usepackage{amsmath}
\usepackage{float}
\usepackage{subfig}
\usepackage{tabularx}
\usepackage[normalem]{ulem}
\usepackage{color}

\begin{document}
\title{Wavelength dependence of reversible photodegradation of disperse orange 11 dye-doped PMMA thin films.}
\author{Benjamin R. Anderson}
\email{branderson@wsu.edu}
%\email{Current Address: Applied Sciences Laboratory, Institute of Shock Physics, Washington State University,
%Spokane, WA 99210-1495}
%\affiliation{Applied Sciences Laboratory, Institute of Shock Physics, Washington State University,
%Spokane, WA 99210-1695}
\affiliation{Department of Physics and Astronomy, Washington State University,
Pullman, WA 99164-2814}
\author{Sheng-Ting Hung}
\affiliation{Department of Physics and Astronomy, Washington State University,
Pullman, WA 99164-2814}
\author{Mark G. Kuzyk}
\affiliation{Department of Physics and Astronomy, Washington State University,
Pullman, WA 99164-2814}
\date{\today}

\begin{abstract}
Using transmittance imaging microscopy we measure the wavelength dependence of reversible photodegradation in disperse orange 11 (DO11) dye-doped (poly)methyl-methacrylate (PMMA).  The reversible and irreversible inverse quantum efficiencies (IQEs) are found to be constant over the spectral region investigated, with the average reversible IQE being $\overline{B}_\alpha= 8.70 (\pm 0.38)\times 10^5$ and the average irreversible IQE being $\overline{B}_\epsilon= 1.396 (\pm 0.031)\times 10^8$.  The large difference between the IQEs is hypothesized to be due to the reversible decay channel being a direct decay mechanism of the dye, while the irreversible decay channel is an indirect mechanism, with the dye first absorbing light, then heating the surrounding environment causing polymer chain scission and cross linking. Additionally, the DO11/PMMA's irreversible IQE is found to be among the largest of those reported for organic dyes, implying that the system is highly photostable. We also find that the recovery 
rate is independent of wavelength, with a value of $\overline{\beta}=3.88(\pm 0.47) \times 10^{-3}$ min$^{-1}$.  These results are consistent with the correlated chromophore domain model of reversible photodegradation.

\vspace{1em}
OCIS Codes: (140.3330) Laser Damage; (140.3380) Laser Materials; (160.4890) Organic Materials; (160.5470) Polymers; (310.0310) Thin Films

\end{abstract}

\maketitle

\vspace{1em}

\section{Introduction}
Photodegradation of organic dye-doped polymers has been a topic of intense study for several decades \cite{wood03.01,taylo05.01,Avnir84.01,Knobbe90.01,Kaminow72.01,Rabek89.01,Gonzalez00.02,Gonzalez00.01,Gonzalez99.01,Gonzalez01.01,Gonzalez03.01,Rezzonico07.01,Zhang98.01} with the majority of dye-doped polymers found to irreversibly photodegrade under intense illumination.  However, in 1998 Peng \textit{et al} reported the first observation of self-healing after photodegradation (i.e. reversible photodegradation) in Rhodamine B and Pyrromethene dye-doped (poly)methyl-methacrylate (PMMA) \cite{Peng98.01}.  Since Peng \textit{et al.}'s study, self-healing has been observed in disperse orange 11 (DO11) dye-doped PMMA \cite{howel04.01,howel02.01}, DO11 dye-doped copolymer of styrene and MMA \cite{Hung12.01},  anthraquinone-derivative-doped PMMA \cite{Anderson11.02}, 8-hydroxyquinoline (Alq) dye-doped PMMA \cite{Kobrin04.01}, air force 455 (AF455) dye-doped PMMA \cite{Zhu07.01}, and Rhodamine 6G/ZrO$_2$ doped 
polyurethane \cite{Anderson15.01}.  While self-healing has been observed in many different materials, the majority of research has focused on DO11/PMMA \cite{howel02.01,howel04.01,embaye08.01,Ramini12.01,Ramini13.01,embaye08.01,Anderson14.02, Anderson11.01, Anderson13.01,Anderson14.03, raminithesis,andersonthesis, Anderson14.01}.

Given the large amount of research focused on DO11/PMMA, it has become the testbed system for studying reversible photodegradation with various probe methods used, such as: amplified spontaneous emission (ASE) \cite{howel02.01,howel04.01,embaye08.01,Ramini12.01,Ramini13.01}, absorption  \cite{embaye08.01,Anderson14.02}, transmittance microscopy \cite{Anderson11.01, Anderson13.01}, white light interferometry \cite{Anderson14.03}, and fluorescence \cite{raminithesis}.  DO11/PMMA's decay and recovery properties have been characterized for differing intensities \cite{Anderson11.01,Anderson14.01,Anderson14.02}, concentrations  \cite{Ramini12.01,Ramini13.01,raminithesis}, temperatures \cite{Ramini12.01,Ramini13.01,raminithesis,andersonthesis}, applied electric fields \cite{Anderson13.01,andersonthesis,Anderson14.01}, and co-polymer compositions \cite{Hung12.01}.

From all these experiments, a model -- based on domains of correlated chromophores -- has been developed, which is found to describe the trends of all the experimental data under the range of conditions studied.  This model is known as the correlated chromophore domain model (CCDM) \cite{Ramini13.01,Anderson14.01,Anderson14.02,raminithesis,andersonthesis}. One of the untested assumptions of the CCDM is that both the reversible and irreversible decay channels are functions of only intensity and independent of the pump wavelength. If this assumption is violated, it would hint at a new unexplored physical mechanism underlying DO11/PMMA's photodegradation.

Therefore in this study we measure reversible photodegradation of DO11/PMMA at five different wavelengths and determine the degradation quantum efficiencies and the recovery rate as a function of pump-photon energy to test if the degradation quantum efficiencies and the recovery rate are dependent on the pump wavelength.  Additionally, we make comparisons of DO11/PMMA's degradation quantum efficiencies to previous studies of {\em irreversible} photodegradation in other dye-doped systems
\cite{Gonzalez00.02,Gonzalez00.01,Gonzalez99.01,Gonzalez01.01,Gonzalez03.01,Rezzonico07.01,Zhang98.01} to show that DO11 compares favorably with the most damage-resistant molecules.

\section{Theory}
To model experimental observations of reversible photodegradation we use a phenomenological three-species population model in which undamaged chromophores (population number density $n_0$) decay either reversibly to species \#1 (of number density $n_1 (z,t)$ at depth $z$ and time $t$) or irreversibly to species \#2 (of number density $n_2 (z,t)$ at depth $z$ and time $t$) taking into account pump depletion as a function of depth, $z$. Figure \ref{fig:model} shows a schematic diagram of the population flows as marked with arrows.  Note that this model is a simplification of the model described in detail in Ref.\cite{Anderson14.01,Anderson14.02} and that for the purposes of this study we do not average over the distribution of domain sizes with the understanding that the parameters determined in this manner are average values. The population dynamics and intensity depletion are described by four coupled differential equations:

\begin{figure}
\centering
\includegraphics{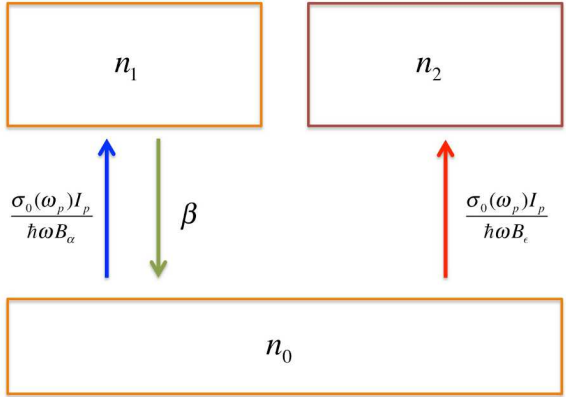}
\caption{Schematic diagram of the three-species population model.  Note that the pump intensity depends on depth and time.}
\label{fig:model}
\end{figure}

\begin{align}
\frac{\partial n_0(z,t)}{\partial t}&=-\frac{\sigma_0(\omega_p)I_p(z,t)}{\hbar\omega}\left(\frac{1}{B_\alpha}+\frac{1}{B_\epsilon}\right) n_0(z,t) \nonumber \\
& + \beta n_1(z,t), \label{eqn:n0}
\\ \frac{\partial n_1(z,t)}{\partial t} &= \frac{\sigma_0(\omega_p)I_p(z,t)}{B_\alpha \hbar\omega_p} n_0(z,t)-\beta n_1(z,t),
\\ \frac{\partial n_2(z,t)}{\partial t}&=\frac{\sigma_0(\omega_p)I_p(z,t)}{B_\epsilon \hbar\omega_p} n_0(z,t), \label{eqn:n2}
\\ \frac{\partial I_p(z,t)}{\partial z}&=-[\sigma_0(\omega_p)n_0(z,t)+\sigma_1(\omega_p)n_1(z,t)\nonumber
\\ &+\sigma_2(\omega_p)n_2(z,t)]I_p(z,t).\label{eqn:Id}
\end{align}
where $\hbar\omega_p$ is the pump photon energy, $B_\alpha^{-1}$ is the quantum efficiency (QE) of reversible degradation, $B_\epsilon^{-1}$ is the QE of irreversible degradation, $I_p(z,t)$ is the pump intensity in W/cm$^2$ at time $t$ and depth $z$, $\beta$ is the recovery rate, and $\sigma_i(\omega_p)$ is the absorbance cross section of the $i^{th}$ species at the pump frequency.

While Equations \ref{eqn:n0}-\ref{eqn:n2} describe the populations as a function of depth and time during photodegradation and recovery, what we physically measure is a change in absorbance, $\Delta A(t;\omega)$, which is related to the cross-sections via,
\begin{equation}
\Delta A(t;\omega)=\Delta \sigma_1(\omega)\int_0^L n_1 (z,t) dz+\Delta \sigma_2(\omega) \int_0^L n_2 (z,t) dz\label{eqn:absdiff}
\end{equation}
where $\Delta\sigma_i=\sigma_i(\omega)-\sigma_0(\omega)$ and $L$ is the thickness of the sample.  Equation \ref{eqn:Id} can be re-expressed in terms of $\Delta\sigma_i$, yielding
\begin{align}
\frac{\partial I_p(z,t)}{\partial z}  = & - [\sigma_0(\omega_p) N + \Delta \sigma_1(\omega_p)n_1(z,t) \nonumber
 \\
 & + \Delta \sigma_2(\omega_p)n_2(z,t)]I_p(z,t).\label{eqn:Id2}
\end{align}
where $N = n_0 + n_1 +n_2$ is the total number density, which in the absence of mass transport is a constant by conservation of the total number of molecules. For the 9 g/l samples used in this study the total dye number density is $N = 2.28 \times 10^{25}$ m$^{-3}$.

Equations \ref{eqn:n0} to \ref{eqn:n2}, Equation \ref{eqn:absdiff} and Equation \ref{eqn:Id2} can be numerically integrated to determine the adjustable parameters $B_\alpha$, $B_\epsilon$, $\Delta\sigma_1$, $\Delta\sigma_2$ and $\beta$ using only a measurement of the observed change in absorbance as a function of time.

At the start of the experiment, the sample is prepared with a known dye concentration consisting only of undamaged species, so $n_0(z,0) \equiv N$ and $n_1 (z,0) = n_2 (z,0) = 0$.  The absorption spectrum is measured at $t=0$ to determine $\sigma_0 (\omega)$, which spans the pump cross-section $\sigma_0 (\omega_p)$.  With these $t=0$ values, the initial depth-dependence of the intensity of the pump beam is obtained by integrating Equation \ref{eqn:Id2}, which yields $I_p(z,0) = I_p \exp[-\sigma_0(\omega_p)z]$, where  $I_p$ is the incident pump intensity.  Given these initial conditions, an iterative numerical integration technique can be applied to solve the coupled differential equations, as follows.

At $t=0$, all of the quantities on the right-hand side of Equations \ref{eqn:n0}-\ref{eqn:n2} are known (i.e. $I_p(z,0)$, $n_0 (z, 0)$, $n_1(z, 0)$ and $n_2(z, 0)$ ) or are adjustable parameters, so the populations at a small time interval later, $t = \Delta t$, can be determined numerically.  Given the updated values $n_0 (z, \Delta t)$, $n_1(z, \Delta t)$ and $n_2(z, \Delta t)$, Equation \ref{eqn:Id2} can be numerically integrated to get $I_p(z, \Delta t)$.  The process is then iterated with the updated depth-dependence of the pump intensity and populations at $t = \Delta t$ to calculate the updated populations and intensity at $t = 2 \Delta t$, and so on.  As a result, the time and depth dependence of the populations and intensity are determined as a function of the fit parameters.

Once the populations as a function of time and depth are determined from the above iterative technique, they can be used to determine the change in the measured absorbance at the probe wavelength as a function of time using Equation \ref{eqn:absdiff}.  A fit of the time dependence of the absorbance to Equation \ref{eqn:absdiff} determines the adjustable parameters $B_\alpha$, $B_\epsilon$, $\Delta\sigma_1$, $\Delta\sigma_2$ and $\beta$.

To avoid confusion, we have calculated only one ray that propagates perpendicular to the plane of the sample and omitted mention of the fact that the intensity depends on the transverse position from the beam center. A beam with a transverse intensity profile, such as a Gaussian beam that is not strongly diverging in the sample at its focal plane, can be viewed as a collection of parallel rays -- each with its own incident intensity.  As such, the incident intensity depends on the transverse coordinates, or $I_p = I_p (x,y)$.  A CCD detector can thus be used to measure the collection of rays, where each pixel selects one such ray with its corresponding intensity. Therefore we can fit the measured change in absorbance at different positions to obtain the change in absorbance as a function of intensity.  We use over 1000 different positions on the sample for each decay and recovery run, which provides ample statistics to determine the values of the five adjustable parameters.

In our experiments, an imaging system is used to first measure the beam profile without the sample, then the experiments are performed with the sample in place to determine the change of absorbance for each ray within the beam.  Since each ray corresponds to a fixed incident intensity, the imaging apparatus can simultaneously determine the change of absorbance for a broad range of input intensities.  As such, a large set of data is generated for data fitting, making it possible to extract the values of the adjustable parameters with a high degree of robustness.

The fit parameters determined from the above procedure are used to compare DO11/PMMA's photodegradation dynamics with other dye-doped systems by using the common figure of merit (FoM) defined by \cite{Gonzalez00.02}:
\begin{align}
\text{FoM}_{R}=\frac{B_\alpha}{\sigma_0},
\end{align}
and
\begin{align}
\text{FoM}_{IR}=\frac{B_\epsilon}{\sigma_0},
\end{align}
where FoM$_R$ is the reversible figure of merit, and FoM$_{IR}$ is the irreversible figure of merit.  The FoM represents the photostability of a material, with larger FoM corresponding to greater resistance to photodegradation.

It is important to emphasize that this model is phenomenological and fits the experimental data well, \cite{Ramini12.01,Ramini13.01,Anderson14.01,Anderson14.02,raminithesis,andersonthesis} but does not reveal the underlying mechanisms of the reversible and irreversible photodegradation pathways, which remains a topic of active study.  Recently reversible photodegradation studies that measure the influence of an applied electric field have suggested that the reversible damage pathway is related to the formation of charged or polarizable fragments that recombine during recovery \cite{Anderson13.01, andersonthesis, Anderson14.01}. This suggests that the reversible QE, $B_\alpha^{-1}$, is a direct measure of the dissociation probability associated with the charged fragments.  With regards to the irreversible damage pathway, recent measurements of the complex index of refraction during decay and recovery suggest that the irreversibly damaged species is related to the formation of damaged polymer-dye complexes, 
with the polymer damage most likely being due to chain scission and cross linking \cite{Anderson14.03}. Given that the polymer is transparent at the pump wavelength, it has been hypothesized that the damage to the polymer is thermally driven by the dye molecule's absorption of light followed by a transfer of heat to the polymer\cite{Anderson14.02,Anderson14.03}.  This hypothesis suggests that the irreversible QE, $B_\epsilon^{-1}$, is a direct measure of the efficiency of thermalization and of the efficiency of thermally induced chain scission and cross-linking \cite{Rabek95.01,Pielichowski05.01,Ichikawa03.01}.

\section{Method}

All experiments presented here use DO11 dye doped in PMMA, with a concentration of 9 g/l, or a molecular number density of $N = 2.28 \times 10^{25}$ m$^{-3}$.  The dye-doped polymer is prepared by first dissolving an appropriate amount of dye into MMA monomer, which is sonicated to ensure that the dye is fully dissolved into solution.  At this point an initiator (butanethiol) and chain transfer agent (Tert-butyl peroxide) are added in the amounts of 33 $\mu$l per 10 ml of MMA, and the solution is then sonicated for 30-60 min.  After sonication the solution is filtered through 0.2 $\mu$m disk filters into vials, which are then placed in a 95 $^\circ$C oven for 48 hr at which point they are fully polymerized.  The vials are then placed in a freezer leading to the solid dye-doped polymer and vial separating due to differential thermal contraction.  Once the solid dye-doped polymer is prepared, a small portion is broken off and placed between two glass slides and thermally pressed at a temperature of 150 $^\circ$C and a uniaxial stress of 90 psi for one hour, allowing the polymer melt to uniformly flow from the center.  Finally, the stress is removed and the sample is allowed to cool.

To measure the dependence of reversible photodegradation on the pump wavelength, we use a transmittance imaging microscope (TIM) \cite{Anderson11.01,Anderson11.02} and an Ar:Kr Ion CW laser pump.  The probe light from a diode source is centered at a wavelength of about $460$ nm and has a spectral width of about $75$ nm.  As such, the CCD probes a range of wavelengths rather than a single one as given by Equation \ref{eqn:absdiff}.  As such, Equation \ref{eqn:absdiff} must be modified by integrating over the intensity spectrum of the beam and the responsivity of the CCD detector.  Assuming that these two functions together give a modification of the form $f_{\omega^\prime}(\omega)$, which peaks at $\omega^\prime$, we can define,
\begin{align}\label{eq:convolute}
\Delta \bar{A}(t;\omega^\prime) = \int_0^\infty d \omega f_{\omega^\prime}(\omega) \Delta A(t;\omega)
\end{align}
and
\begin{align}\label{eq:convolute}
\Delta \bar{\sigma}_i (\omega^\prime) = \int_0^\infty d \omega f_{\omega^\prime}(\omega) \Delta \sigma(\omega) ,
\end{align}
which when applied to Equation \ref{eqn:absdiff} gives,
\begin{align}
\Delta \bar{A}(t;\omega^\prime )= \Delta \bar{\sigma}_1(\omega^\prime)\int_0^L n_1 (z,t) dz \nonumber
\\ +\Delta \bar{\sigma}_2(\omega^\prime) \int_0^L n_2 (z,t) dz. \label{eqn:absdiff2}
\end{align}
As such, the analysis is the same a for the single-frequency case with the understanding that all the quantities are barred.

The power exiting the laser head used to pump the sample at 476 nm, 488 nm, 496 nm, and 514 nm is maintained at 1 W, with the on-sample power being measured using an Edmund Optics power meter.  We also use the 502 nm and 457 nm lines but with a power exiting the laser of 353 mW and 373 mW, respectively.

\begin{figure}
\centering
\includegraphics{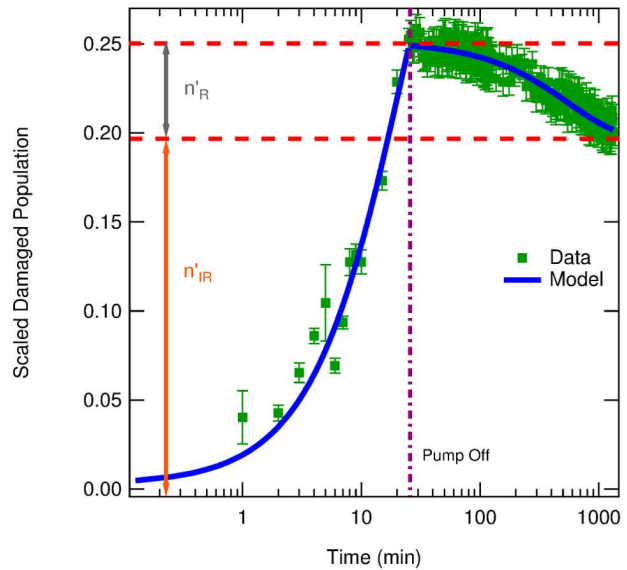}
\caption{Scaled damaged population measured during decay and recovery (points) and a fit to the three level model (curve).  The dashed horizontal lines mark the peak SDP and the SDP after recovery.  The area between the dashed lines corresponds to the reversibly damaged population.}
\label{fig:expdec}
\end{figure}

\begin{figure}
\centering
\includegraphics{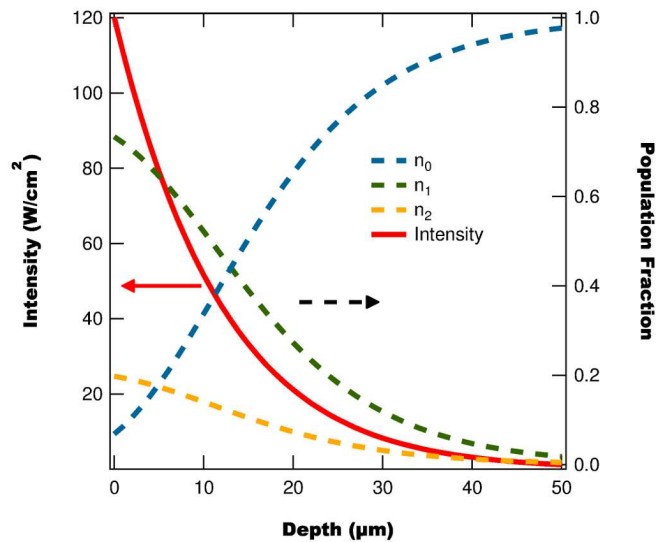}
\caption{Pump Intensity and population fractions as a function of depth at $t=5$ min.  The pump laser is turned on at $t=0$ to initiate photodegradation.}
\label{fig:depthcomp}
\end{figure}

The beam profile for each wavelength is measured using the TIM and neutral density filters, allowing us to correlate the beam intensity profile to the sample's damage profile.  This allows for a precise determination of the damage as a function of intensity.  For the bright laser lines (476 nm, 488 nm, 496 nm, and 514 nm) degradation is initiated for three minutes, while for the dim laser lines (502 nm and 457 nm) the time is extended to ten minutes.  Images are recorded at intervals of thirty seconds during degradation and semi-log intervals during recovery. The images obtained during decay and recovery are a mapping of intensity for a given $x$ and $y$ position, $C(x,y)$, which is converted to the scaled damaged population (SDP) \cite{Anderson11.01,Anderson11.02} using,

\begin{align}
n'(x,y,t)&=-\ln \left(\frac{C(x,y,t)}{C_0}\right)
\\ &= \Delta \bar{A}(x,y,t),
\end{align}
where $C_0$ is the background pixel intensity and $\Delta \bar{A}(x,y,t)$ is the change in absorbance at point $x,y$.  The SDP is then fit to Equation \ref{eqn:absdiff2} to determine the decay and recovery parameters.  Figure \ref{fig:expdec} shows an example decay and recovery curve as a function of time with the reversible and irreversible contributions labelled. We can also plot the intensity and fractional populations as a function of depth, shown in Figure \ref{fig:depthcomp}, at $t=5$ min, determined from fitting the model to experimental results. From Figure \ref{fig:depthcomp} we see that the majority of the damaged species is found near the surface of the sample and decays with the functional form of a stretched exponential with depth into the sample, while the fraction of undamaged molecules increases with depth.  We also find that the pump intensity as a function of depth follows a double exponential, which makes sense given that the absorbance cross sections of the damaged and undamaged species 
are different.  By measuring the populations during decay and recovery over a wide range of intensities, we can determine the reversible FoM, the irreversible FoM, and the recovery rate. Note that the use of a large number of decay and recovery curves, approximately 1000  of them, enables all the fit parameters to be determined to a high degree of accuracy.  In other words, while the parameters for a single intensity decay and recovery curve has a large degree of uncertainty due to the model's complexity, fitting all the data sets simultaneously gives better statistics that greatly improve the accuracy of the fit parameters.

\section{Results and Discussion}
\begin{figure}
\centering
\includegraphics{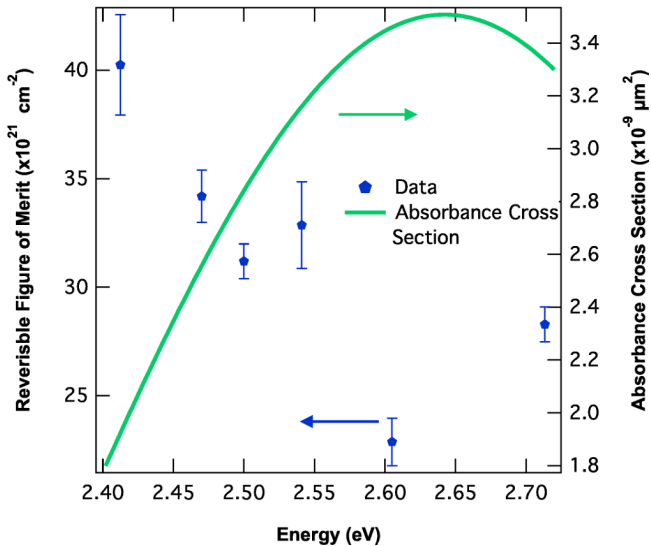}
\caption{Figure of merit for the reversible decay process as a function of pump photon energy.}
\label{fig:alp}
\end{figure}

\begin{figure}
\centering
\includegraphics{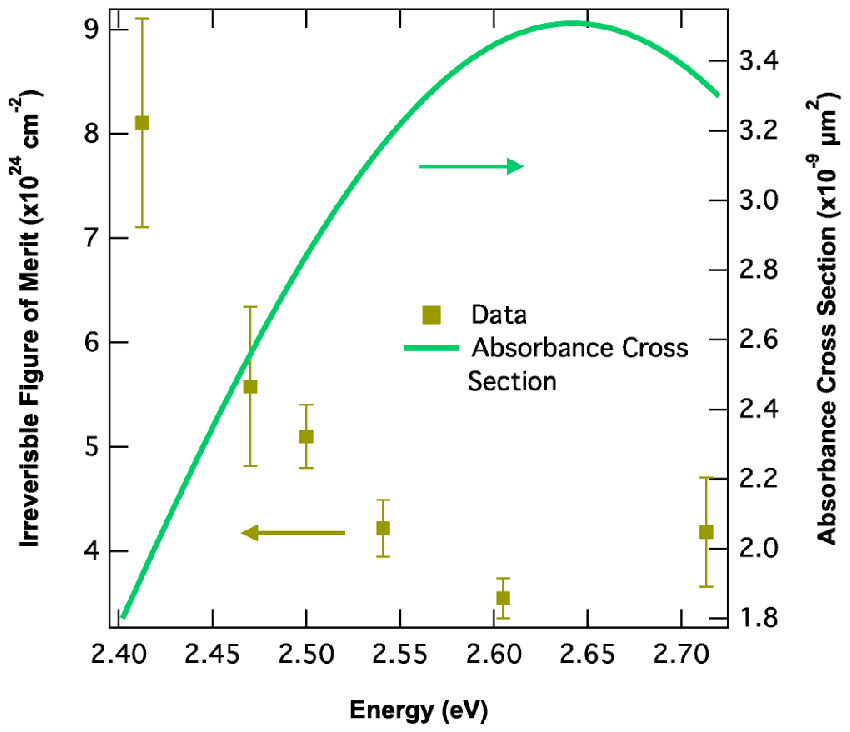}
\caption{Figure of merit for the irreversible decay process as a function of pump photon energy.}
\label{fig:eps}
\end{figure}

From imaging measurements during photodegradation we determine the reversible and irreversible FoMs. Figure \ref{fig:alp} shows the reversible FoM and Figure \ref{fig:eps}, the irreversible FoM.  Both are found to scale inversely with the undamaged absorbance cross section, which implies that the inverse quantum efficiency (IQE) is constant over the photon energies tested, consistent with results for other dye-doped polymers \cite{Gonzalez00.02,Gonzalez00.01,Gonzalez99.01,Gonzalez01.01,Gonzalez03.01,Rezzonico07.01,Zhang98.01}.

\begin{table}
\begin{center}
\begin{tabular}{|c|cc|}
\hline
\textbf{Energy (eV)}   &  \textbf{B$_\alpha$ (10$^{6}$)}  & \textbf{B$_\epsilon$ (10$^{8}$)}   \\ \hline
2.41   &   $0.70  \pm  0.20$  &  $1.40  \pm  0.18$  \\[0.5em]
2.47   &   $0.85  \pm  0.15$  &  $1.383  \pm  0.095 $ \\[0.5em]
2.50   &   $0.877 \pm  0.051$  &  $1.434 \pm 0.043$ \\[0.5em]
2.54   &   $1.04 \pm 0.14$   &  $1.329 \pm 0.086$ \\[0.5em]
2.60   &   $0.797 \pm 0.090$   &  $1.28 \pm 0.13$ \\[0.5em]
2.71   &   $0.94 \pm 0.13$  &  $1.386 \pm 0.043$ \\
\hline
\end{tabular}
\end{center}
\caption{Reversible, $B_{\alpha}$, and irreversible, $B_\epsilon$, inverse quantum efficiency for different pump energies. The irreversible inverse quantum efficiency is $\approx 160\times$ larger than the reversible inverse quantum efficiency.}
\label{Tab:fits}
\end{table}

\begin{figure}
\centering
\includegraphics{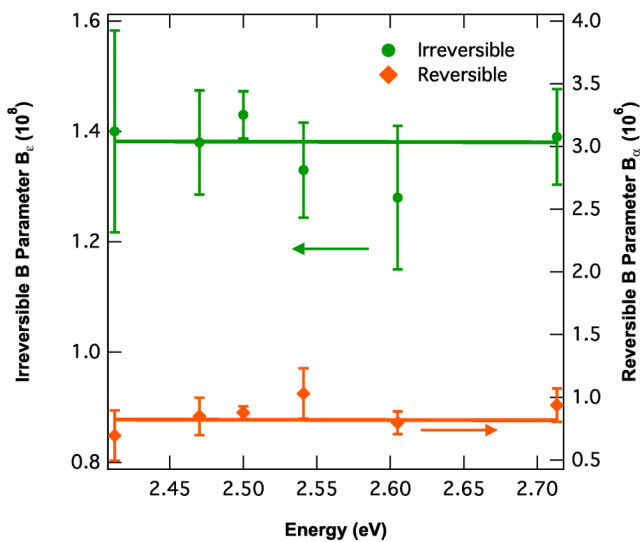}
\caption{Reversible, $B_{\alpha}$, and irreversible, $B_\epsilon$, inverse quantum efficiency as a function of pump photon energy.}
\label{fig:Bcomp}
\end{figure}

Multiplying the FoMs by the undamaged absorbance cross section we determine the reversible and irreversible IQEs, which are tabulated in Table \ref{Tab:fits} and displayed as a function of pump energy in Figure \ref{fig:Bcomp}.  Both the reversible and irreversible IQE are found to be constant, within experimental uncertainty,  as a function of photon pump energy. The weighted average IQEs are $\overline{B}_\alpha= 8.70( \pm 0.38)\times 10^5$ and $\overline{B}_\epsilon= 1.396 (\pm 0.031)\times 10^8$.  The irreversible IQE is $\approx 160$ times larger than the reversible IQE, so DO11/PMMA has a greater probability of reversible degradation than irreversible.  This result is consistent with previous observations of the relative magnitudes between the two in DO11/PMMA  \cite{Anderson14.01,Anderson14.02,Anderson14.03}.  In addition to comparing the IQEs of the two degradation pathways we also
compare DO11/PMMA's IQEs to those of other dye-doped polymer systems \cite{Dubois96.01,Gonzalez00.01,Gonzalez00.02,Gonzalez01.01}.  In other dye doped systems the {\em irreversible} IQE is typically on the order of $10^6$  \cite{Gonzalez00.02,Gonzalez00.01,Gonzalez99.01,Gonzalez01.01,Gonzalez03.01,Rezzonico07.01,Zhang98.01}, which is similar to DO11/PMMA's {\em reversible} IQE, but still two orders of magnitude smaller than DO11/PMMA's irreversible IQE.  A review of the literature reveals only a handful of other dyes with IQE's greater than $5 \times 10^7$, which are tabulated in Table \ref{tab:litrev}. This result implies that  DO11/PMMA is among the most resistant dyes to irreversible photodegradation.

\begin{table*}
\begin{tabular}{|>{\centering\arraybackslash}m{2cm}>{\centering\arraybackslash}m{3cm}>{\centering\arraybackslash}m{3cm}>{\centering\arraybackslash}m{1.2cm}|}
\hline
\textbf{Dye Name}  &  \textbf{Structure}  &  \textbf{B$_\epsilon$}   &  \textbf{Ref.} \\ \hline
DO11  &   \parbox[c]{1em}{\includegraphics{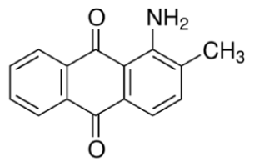}}  &  $1.4\times 10^8$   &   -- \\ \hline
Rhodamine B  &  \parbox[c]{1em}{\includegraphics{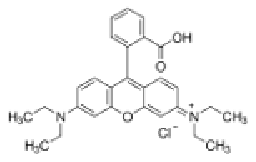}}  & $8\times10^7$ &  \cite{Dubois96.01} \\ \hline
Perylene Red &  \parbox[c]{1em}{\includegraphics{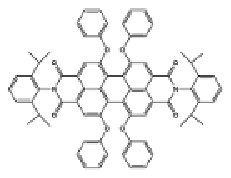}}  &$9\times10^8$  &  \cite{Dubois96.01} \\ \hline
Perylene Orange  &  \parbox[c]{1em}{\includegraphics{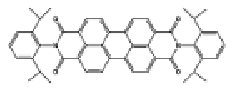}}  &$1\times10^9$  &  \cite{Dubois96.01} \\ \hline
-  &  \parbox[c]{1em}{\includegraphics{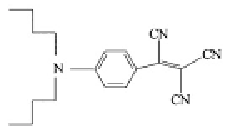}}  &$1\times 10^8$  &  \cite{Gonzalez00.01} \\ \hline
-  &  \parbox[c]{1em}{\includegraphics{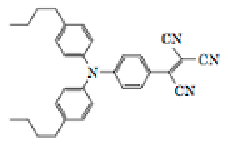}}  &$5\times 10^7$  &  \cite{Gonzalez01.01} \\ \hline
-  &  \parbox[c]{1em}{\includegraphics{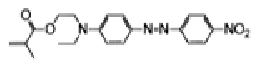}}  &$3\times 10^8$  &  \cite{Gonzalez00.02} \\ \hline
-  &  \parbox[c]{1em}{\includegraphics{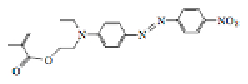}}  &$1\times 10^8$  &  \cite{Gonzalez01.01} \\ \hline
\end{tabular}
\caption{Tabulation of some of the most photodamage resistant dyes reported in the literature and their irreversible IQE.  The majority of dyes have IQEs on the order of $10^6$.}
\label{tab:litrev}
\end{table*}

One possible explanation for DO11/PMMA's greater resistance to irreversible degradation is based on our recent hypothesis that the reversibly damaged species is related to a photo-induced reversible change in the dye molecule, while the the irreversibly damaged species is due to the formation of a damaged dye-polymer complex \cite{Anderson14.02,Anderson14.03}.  Since neat PMMA is transparent in the visible, photodamage to the polymer must be mediated by energy transfer between the dye molecules and the polymer \cite{Moshrefzadeh93.01,Gonzalez00.01,Chang01.01,Annieta,Rabek95.01,Fellows05.01,Ishchenko08.01}.  This implies an indirect damage mechanism for the irreversibly damaged species, which is less efficient than a direct damage mechanism. Since DO11 is directly damaged in the spectral region of the pump wavelengths used, the reversibly damaged species will form more efficiently than the irreversibly damaged species, resulting in the observed relationship between the IQEs.

\begin{figure}
\centering
\includegraphics{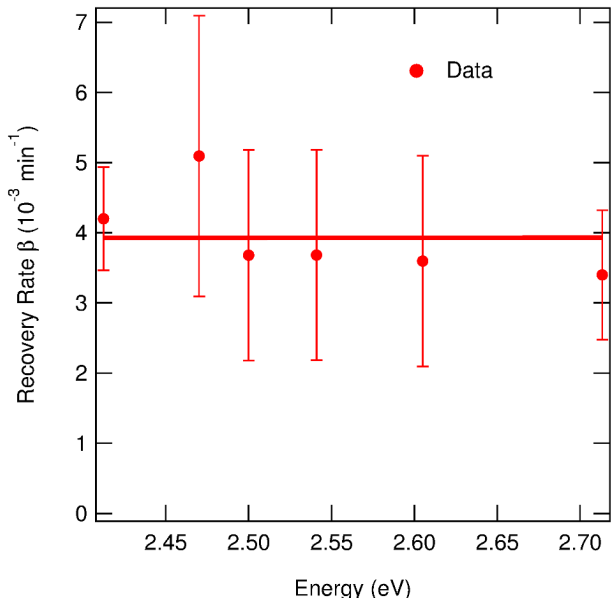}
\caption{Recovery rate as a function of pump-photon energy.  All recovery rates are found to be within experimental uncertainty of each other.}
\label{fig:bet}
\end{figure}

After degradation, we measure recovery dynamics of DO11/PMMA as a function of pump-photon energy.  Figure \ref{fig:bet} shows the recovery rate as a function of pump-photon energies, which is found to be constant within experimental uncertainty for the spectral region used. The average recovery rate over the spectral region is $\overline{\beta}=3.88(\pm 0.47) \times 10^{-3}$ min$^{-1}$, which is consistent with previous measurements of 9 g/l DO11/PMMA \cite{embaye08.01,Ramini13.01,Anderson11.01,Anderson13.01, Anderson14.01,Anderson14.02}.

The observation of the decay parameters, $B_\alpha$ and $B_\epsilon$, and recovery rate, $\beta$, being independent of pump wavlength wavelength in the spectral region used is consistent with the assumptions of the CCDM \cite{Ramini13.01,Anderson14.01}.  While this result by itself does not prove the CCDM to be the correct model of reversible photodegradation, the observation of wavelength-independent reversible photodegradation is additional evidence supporting its validity.

Finally, the fit parameters also give the change in the average cross-section over the spectral range of the probe beam during a decay run, which yields $\Delta \bar{\sigma}_1 / \Delta \bar{\sigma}_2 = 2.88( \pm 0.34) \times 10^{-2}$.  As such, the change in the optical absorption cross section of the reversibly-damaged species within the probe spectral range is much smaller than the change of the cross section of the irreversibly-damaged species.  This implies that the change in conjugated bond length due to reversible degradation is much smaller than the change due to irreversible damage.
 However, the reader should be warned that a large dispersion in the sensitivity of the detector could somewhat change the value of this ratio, but the effect is most likely small.

\section{Conclusion}

Studies of reversible photodegradation in DO11/PMMA have been ongoing for over a decade with much progress made in shedding light on the underlying mechanisms and relevant parameters.  However, these studies have primarily used a single pump wavelength, without consideration of how self-healing depends on it.  The wavelength dependence of DO11/PMMA's degradation and recovery can help probe particular mechanisms as well as eliminate others. We therefore have used various Ar:Kr laser lines (457 nm, 476 nm, 488 nm, 496 nm, 502 nm, 514 nm) to determine the relevant decay and recovery parameters as a function of wavelength using a simplified version of the depth dependent eCCDM, which describes two different decay pathways -- one reversible and the other irreversible \cite{Anderson14.01,Anderson14.02,andersonthesis}.

By fitting the experimental data to the model we determine the fundamental photodegradation properties of DO11/PMMA  ($B_\alpha$, $B_\epsilon$, and $\beta$), which are found to be constant as a function of pump wavelength within the spectral region investigated. The decay and recovery parameters are found to be independent of wavelength, which is consistent with previous measurements of irreversible photodegradation in other dye-doped systems \cite{Gonzalez00.02,Gonzalez00.01,Gonzalez99.01,Gonzalez01.01,Gonzalez03.01,Rezzonico07.01,Zhang98.01}, as well as with the assumptions of the correlated chromophore domain model \cite{Ramini13.01,Anderson14.01}.

We also find that DO11/PMMA is highly resistant to irreversible photodegradation, where its irreversible IQE is comparable to the most damage resistant dyes reported in literature; making it a promising candidate for high intensity optical applications. Additionally, the irreversible IQE is two orders of magnitude higher than the reversible IQE.  This observation is consistent with the hypothesis that the irreversible species is formed by photothermally driven chain scission and cross linking of the polymer, with local photothermal heating being provided by the dye molecules \cite{Anderson14.03}. In this scheme the reversibly decayed species is formed directly by the interaction of the dye and light; while the irreversible species is formed by an indirect process requiring the dye to first absorb light, then transferring the energy as heat to the local environment, which degrades the polymer.  This indirect process is less efficient than direct interaction, which is borne out out by the data.

The wavelength dependence determined here is consistent with the assumptions of the domain model, and provides another piece to the puzzle of the mechanisms underlying reversible photodegradation.  The fact that the direct damage process is governed by the absorption of a photon, which is characterized by the linear absorption cross section, pegs it as a linear process that sometimes produces a damaged fragment.  As such, nonlinear processes are ruled out.  Furthermore, since the time constant of the recovery process is independent of the wavelength, the species produced is the same for each wavelength.  Future work will extend the measurement domain to temperature and concentration dependence to to test the model's predictions beyond the range in which it was developed.

\section{Acknowledgments}
We thank Wright Patterson Air Force Base and Air Force Office of Scientific Research (FA9550- 10-1-0286) for theirsupport of this research.

%\bibliographystyle{osajnl}
%\bibliography{PrimaryDatabase,ASLbib}

\newpage

\begin{thebibliography}{10}
\newcommand{\enquote}[1]{``#1''}

\bibitem{wood03.01}
R.~M. Wood, \emph{Laser-Induced Damage of Optical Materials}, Series in Optics
  and Optoelectronics (Taylor \& Francis, Boca Raton, 2003).

\bibitem{taylo05.01}
E.~W. Taylor, J.~E. Nichter, F.~D. Nash, F.~Haas, A.~A. Szep, R.~J. Michalak,
  B.~M. Flusche, P.~R. Cook, T.~A. McEwen, B.~F. McKeon, P.~M. Payson, G.~A.
  Brost, A.~R. Pirich, C.~Castaneda, B.~Tsap, and H.~R. Fetterman,
  \enquote{{Radiation resistance of electro-optic polymer-based modulators},}
  Appl. Phys. Lett. \textbf{86}, 201122 (2005).

\bibitem{Avnir84.01}
D.~Avnir, D.~Levy, and R.~Reisfeld, \enquote{The nature of the silica cage as
  reflected by spectral changes and enhanced photostability of trapped
  rhodamine 6g,} J. Phys. Chem \textbf{88}, 5956--5959 (1984).

\bibitem{Knobbe90.01}
E.~T. Knobbe, B.~Dunn, P.~D. Fuqua, and F.~Nishida, \enquote{Laser behavior and
  photostability characteristics of organic dye doped silicate gel materials,}
  Appl. Opt \textbf{29}, 2729--2733 (1990).

\bibitem{Kaminow72.01}
I.~P. Kaminow, L.~W. Stulz, E.~A. Chandross, and C.~A. Pryde,
  \enquote{Photobleaching of organic laser dyes in solid matrices,} Appl. Opt
  \textbf{11}, 1563--1567 (1972).

\bibitem{Rabek89.01}
J.~F. Rabek and J.-P. Fouassier, \emph{Lasers in Polymer Science and
  Technology: Applications, Volume I} (CRC Press, 1989).

\bibitem{Gonzalez00.02}
A.~Galvan-Gonzalez, M.~Canva, G.~I. Stegeman, L.~Sukhomlinova, R.~J. Twieg,
  K.~P. Chan, T.~C. Kowalczyk, and H.~S. Lackritz, \enquote{Photodegradation of
  azobenzene nonlinear optical chromophores: the influence of structure and
  environment,} JOSA B \textbf{17}, 1992--2000 (2000).

\bibitem{Gonzalez00.01}
A.~Galvan-Gonzalez, M.~Canva, G.~I. Stegeman, R.~Twieg, K.~P. Chan, T.~C.
  Kowalczyk, X.~Q. Zhang, H.~S. Lackritz, S.~Marder, and S.~Thayumanavan,
  \enquote{Systematic behavior of electro-optic chromophore photostability,}
  Optics Lett. \textbf{25}, 332--334 (2000).

\bibitem{Gonzalez99.01}
A.~Galvan-Gonzalez, M.~Canva, G.~I. Stegeman, R.~Twieg, T.~C. Kowalczyk, and
  H.~S. Lackritz, \enquote{Effect of temperature and atmospheric environment on
  the photodegradation of some disperse red 1-type polymers,} Optics Lett.
  \textbf{24}, 1741--1743 (1999).

\bibitem{Gonzalez01.01}
A.~Galvan-Gonzalez, G.~I. Stegeman, A.~K.-Y. Jen, X.~Wu, M.~Canva, A.~C.
  Kowalczyk, X.~Q. Zhang, H.~S. Lackritz, S.~Marder, S.~Thayumanavan, and
  G.~Levina, \enquote{Photostability of electro-optic polymers possessing
  chromophores with efficient amino donors and cyano-containing acceptors,}
  JOSA B \textbf{18}, 1846--1853 (2001).

\bibitem{Gonzalez03.01}
A.~Galvan-Gonzalez, K.~D. Belfield, G.~I. Stegeman, M.~Canva, S.~R. Marder,
  K.~Staub, G.~Levina, and R.~Tweig, \enquote{Photodegradation of selected
  $pi$-conjugated electro-optic chromophores,} J. Appl. Phys. \textbf{94}, 756
  (2003).

\bibitem{Rezzonico07.01}
D.~Rezzonico, M.~Jazbinsek, P.~Günter, C.~Bosshard, D.~H. Bale, Y.~Liao, L.~R.
  Dalton, and P.~J. Reid, \enquote{Photostability studies of $pi$-conjugated
  chromophores with resonant and nonresonant light excitation for long-life
  polymeric telecommunication devices,} JOSA B \textbf{24}, 2199 (2007).

\bibitem{Zhang98.01}
Q.~Zhang, M.~Canva, and G.~Stegeman, \enquote{Wavelength dependence of
  4-dimethylamino-4-nitrostilbene polymer thin film photodegradation,} App.
  Phys. Lett. \textbf{73}, 912 (1998).

\bibitem{Peng98.01}
G.~D. Peng, Z.~Xiong, and P.~L. Chu, \enquote{{Fluorescence Decay and Recovery
  in Organic Dye-Doped Polymer Optical Fibers},} J. Lightwave Technol.
  \textbf{16}, 2365--2372 (1998).

\bibitem{howel04.01}
B.~Howell and M.~G. Kuzyk, \enquote{{Lasing Action and Photodegradation of
  Disperse Orange 11 Dye in Liquid Solution},} Appl. Phys. Lett. \textbf{85},
  1901--1903 (2004).

\bibitem{howel02.01}
B.~Howell and M.~G. Kuzyk, \enquote{{Amplified Spontaneous Emission and
  Recoverable Photodegradation in Disperse-Orange-11-Doped-Polymer},} J. Opt.
  Soc. Am. B \textbf{19}, 1790 (2002).

\bibitem{Hung12.01}
S.-T. Hung, S.~K. Ramini, D.~G. Wyrick, K.~Clays, and M.~G. Kuzyk, \enquote{The
  role of the polymer host on reversible photodegradation in disperse orange 11
  dye,} in \enquote{SPIE Optics and Photonics: Organic Photonics +
  Electronics,}  (San Diego, CA), 84741A.

\bibitem{Anderson11.02}
B.~R. Anderson, S.~K. Ramini, and M.~G. Kuzyk, \enquote{Imaging studies of
  photodamage and self- healing of anthraquinone derivative dye doped
  polymers.} in \enquote{SPIE Laser Damage Symposium Proc.}, G.~Exarhos, ed.
  (SPIE, Boulder, CO, 2011), [8190-16].

\bibitem{Kobrin04.01}
P.~Kobrin, R.~Fisher, and A.~Gurrola, \enquote{Reversible photodegradation of
  organic light-emitting diodes,} Appl. Phys. Lett. \textbf{85}, 2385 (2004).

\bibitem{Zhu07.01}
Y.~Zhu, J.~Zhou, and M.~G. Kuzyk, \enquote{{Two-photon fluorescence
  measurements of reversible photodegradation in a dye-doped polymer},} Opt.
  Lett. \textbf{32}, 958--960 (2007).

\bibitem{Anderson15.01}
B.~R. Anderson, R.~Gunawidjaja, and H.~Eilers, \enquote{Self-healing random
  lasers,} Optics Letters \textbf{40}, 577--580 (2015).

\bibitem{embaye08.01}
N.~Embaye, S.~K. Ramini, and M.~G. Kuzyk, \enquote{{Mechanisms of reversible
  photodegradation in disperse orange 11 dye doped in PMMA polymer},} J. Chem.
  Phys. \textbf{129}, 054504 (2008).

\bibitem{Ramini12.01}
S.~K. Ramini and M.~G. Kuzyk, \enquote{A self healing model based on
  polymer-mediated chromophore coerrelations,} J. Chem Phys. \textbf{137},
  054705 (2012).

\bibitem{Ramini13.01}
S.~K. Ramini, B.~R. Anderson, S.~T. Hung, and M.~G. Kuzyk,
  \enquote{Experimental tests of a new correlated chromophore domain model of
  self-healing in a dye-doped polymer,} Polymer Chemistry \textbf{4}, 4948
  (2013).

\bibitem{Anderson14.02}
B.~R. Anderson, S.~Hung, and M.~Kuzyk, \enquote{The effect of pump depletion on
  reversible photodegradation.} Opt. Com. \textbf{318}, 180--185 (2014).

\bibitem{Anderson11.01}
B.~R. Anderson, S.~K. Ramini, and M.~G. Kuzyk, \enquote{Imaging studies of
  photodamage and self-healing in disperse orange 11 dye-doped pmma,} J. Opt.
  Soc. Am. B \textbf{28}, 528--32 (2011).

\bibitem{Anderson13.01}
B.~R. Anderson, S.-T. Hung, and M.~G. Kuzyk, \enquote{Electric field dependent
  photodegradation and self healing of disperse orange 11 dye-doped pmma thin
  films,} J. Opt. Soc. Am. B \textbf{30}, 3193--3201 (2013).

\bibitem{Anderson14.03}
B.~R. Anderson and M.~G. Kuzyk, \enquote{Mechanisms of the refractive index change
  in do11/pmma due to photodegradation,} Optical Materials \textbf{36},
  1227--1231 (2014).

\bibitem{raminithesis}
S.~K. Ramini, \enquote{Experimental investigations of a proposed chromophore
  correlation model of self healing of disperse orange 11 doped in poly(methyl
  methacrylate),} Ph.D. thesis, Washington State University (2012).

\bibitem{andersonthesis}
B.~R. Anderson, \enquote{Testing a generalized domain model of photodegradation
  and self-healing using novel optical characterization techniques and the
  effects of an applied electric field,} Ph.D. thesis, Washington State
  University (2013).

\bibitem{Anderson14.01}
B.~R. Anderson and M.~G. Kuzyk, \enquote{Generalizing the correlated
  chromophore domain model of reversible photodegradation to include the
  effects of an applied electric field,} Phys. Rev E \textbf{89}, 032601
  (2014).

\bibitem{Rabek95.01}
J.~F. Rabek, \emph{Polymer Photodegradation: Mechanisms and experimental
  methods} (Springer-Science+Business Media, 1995).

\bibitem{Pielichowski05.01}
K.~Pielichowski and J.~Njuguna, \emph{Thermal Degradation of Polymeric
  Materials} (iSmithers Rapra Publishing, 2005).

\bibitem{Ichikawa03.01}
T.~Ichikawa, K.~i.~Oyama, T.~Kondoh, and H.~Yoshida, \enquote{Efficiency of
  radiation-induced main-chain scission of poly (methyl methacrylate) depends
  on the irradiation temperature because of coexisting monomer,,} Journal of
  Polymer Science Part A: Polymer Chemistry \textbf{32}, 2487--2492 (1994).

\bibitem{Dubois96.01}
A.~Dubois, M.~Canva, A.~Brun, F.~Chaput, and J.-P. Boilot,
  \enquote{Photostability of dye molecules trapped in solid matrices,} App.
  Optics \textbf{35}, 3193--3199 (1996).

\bibitem{Moshrefzadeh93.01}
R.~S. Moshrefzadeh, D.~K. Misemer, M.~D. Radcliffe, C.~V. Francis, and S.~K.
  Mohapatra, \enquote{Nonuniform photobleaching of dyed polymers for optical
  waveguides,} Appl. phys. lett. \textbf{62}, 16--18 (1993).

\bibitem{Chang01.01}
S.-C. Chang, G.~He, F.-C. Chen, T.-F. Guo, and Y.~Yang, \enquote{Degradation
  mechanism of phosphorescent-dye-doped polymer light-emitting diodes,} App.
  Phys. Lett. \textbf{79}, 2088 (2001).

\bibitem{Annieta}
P.~Annieta, L.~Joseph, L.~Irimpan, P.~Radhakrishnan, and V.~Nampoori,
  \enquote{Photosensitivity of laser dye mixtures in polymer matrix  a
  photoacoustic study,} .

\bibitem{Fellows05.01}
C.~Fellows, U.~Tauber, C.~Carvalho, and C.~Carvalhaes, \enquote{Amplified
  spontaneous emission of proton transfer dyes in polymers,} Brazilian Journal
  of Physics \textbf{35}, 933 (2005).

\bibitem{Ishchenko08.01}
A.~A. Ishchenko, \enquote{Photonics and molecular design of dye-doped polymers
  for modern light-sensitive materials,} Pure Appl. Chem. \textbf{80},
  1525--1538 (2008).

\end{thebibliography}

\end{document}